\documentclass{osa-article}
\journal{osajournal}
\articletype{Research Article}
\usepackage{threeparttable} 
\usepackage{subfigure} 
\usepackage{url} 

\begin{document}

\title{Demonstration of wideband metal mesh filters for submillimeter astrophysics using flexible printed circuits}

\author{Shinsuke Uno,\authormark{1,*}
        Tatsuya Takekoshi,\authormark{1}
        Tai Oshima,\authormark{2,3}
        Keisuke Yoshioka,\authormark{2,4}
        Kah Wuy Chin,\authormark{1,2}
        and Kotaro Kohno\authormark{1,5}
}

\address{\authormark{1}Institute of Astronomy, Graduate School of Science, The University of Tokyo, 2-21-1 Osawa, Mitaka, Tokyo 181-0015, Japan\\
\authormark{2}National Astronomical Observatory of Japan, 2-21-1 Osawa, Mitaka, Tokyo 181-8588, Japan\\
\authormark{3}Department of Astronomical Science, The Graduate University of Advanced Studies (SOKENDAI), Mitaka, Tokyo 181-8588, Japan\\
\authormark{4}Graduate School of Informatics and Engineering, The University of Electro-Communications, 1-5-1 Chofugaoka, Chofu, Tokyo 182-8585, Japan\\
\authormark{5}Research Center for Early Universe, School of Science, The University of Tokyo, 7-3-1 Hongo, Bynkyo-ku, Tokyo 113-0033, Japan}%

\email{\authormark{*}uno@ioa.s.u-tokyo.ac.jp}

\begin{abstract}
We developed a wideband quasi-optical band-pass filter covering 170--520~GHz by exploiting the recent advancements in commercially available flexible printed circuit (FPC) fabrication technologies. 
We designed and fabricated a three-layered stack of loaded hexagonal grid metal meshes using a copper pattern with a narrowest linewidth of $50~\mu\mathrm{m}$ on a polyimide substrate. 
The measured frequency pass-band shape was successfully reproduced through a numerical simulation using a set of parameters consistent with the dimensions of the fabricated metal meshes.  
FPC-based metal mesh filters will provide a new pathway toward the on-demand development of millimeter/submillimeter-wave quasi-optical filters at low cost and with a short turnaround time. 
\end{abstract}

\section{Introduction}

Millimeter- and submillimeter-wave (mm/submm) continuum emissions are a unique clue in investigating the historical structural formation of the universe through observations of galaxy clusters using the Sunyaev-Zel'dovich effect (see \cite{Mroczkowski2019} for a recent review) and massive dusty star-forming galaxies by exploiting a negative K-correction \cite{Casey2014}.
Continuum camera systems based on direct photon detector arrays, including transition edge sensors (TESs) and kinetic inductance detectors (KIDs), are important instruments for conducting sensitive, wide-area mm/submm observations.
Quasi-optical frequency selective filters \cite{Goldsmith_1998} \cite{munk2005frequency} are one of the key components of continuum cameras in accurately defining observation bands of interest. 
Highly reliable quasi-optical filters for mm/submm wavelengths have been developed \cite{ADE2006} and widely used in recent astronomical observational instruments \cite{Takekoshi2012}. 
In particular, broadband measurements of the continuum spectrum is crucial in reliably estimating the physical parameters and depicting the evolution of the galaxies and galaxy clusters in the distant universe \cite{deBernardis2012}, and therefore the importance of conducting simultaneous multi-frequency observations is increasing \cite{Takekoshi2012,Oshima2013}.
Some examples of ground-based multi-frequency facilities include the JCMT 15-m telescope with SCUBA-2 \cite{Holland2013}, the IRAM 30-m telescope with NIKA2 \cite{Adam2018}, the South Pole Telescope \cite{Carlstrom2011}, and the Atacama Cosmology Telescope \cite{Swetz2011}. 
Furthermore, the LMT 50-m telescope with TolTEC \cite{Austermann2018} and CCAT-prime \cite{10.1117/12.2314031}, allowing observations of more than two bands simultaneously, are expected to be operational in the near future. 

To promote the efficient use of multi-frequency continuum observations, we are developing a new focal plane array system for observing three atmospheric windows covering the frequency range of 125--295~GHz, implementing on-chip trichroic broadband filters \cite{Suzuki2012onchip} (centered at 150, 220, and 270~GHz) combined with KIDs for each spatial pixel.
Our new focal plane array also requires quasi-optical wideband band-pass filters (BPFs) with steep high-pass and low-pass cutoff characteristics and good transmittance covering the 125--295~GHz range (fractional bandwidth of 0.8) to eliminate out-of-band emissions.
In particular, to exploit multichroic detectors covering three bands simultaneously, the bandwidth of the quasi-optical BPF must be more than two-times wider (a fractional bandwidth of $\sim$1) than the filters for ``conventional'' multi-frequency observational systems, which have been implemented using narrower BPFs (a typical fractional bandwidth of $\sim$0.2--0.3) for each band \cite{ADE2006}.
Multiple prototyping and laboratory evaluations with a reasonably short turnaround are necessary to implement such filters with challenging specifications, but it is not clear whether such on-demand development is feasible based on the current filter fabrication processes. 

Herein, we propose the use of a commercially available flexible printed circuit (FPC) fabrication technology to implement metal mesh filters as mm/submm-wave BPFs. 
The recent and rapid advancements of commercially available FPC technologies have allowed the fabrication of fine metal patterns 
with a narrowest linewidth of $50~\mathrm{\mu m}$, which is required for filter patterns 
within the mm/submm wavelengths, on a large substrate membrane ($30\times30~\mathrm{cm^2}$) 
at low cost and with a short turnaround time.
Popular commercially available FPCs consist of a copper foil and polyimide substrate layers, and such FPCs can also be used in a cryogenic environment because the thermal contraction of polyimide ($\Delta L/L_{293-4K}=0.44\%$) is roughly the same as that of copper ($\Delta L/L_{293-4K}=0.324\%$) \cite{ekin2006}, and therefore a peeling of the copper foil does not occur. 
In fact, it has been demonstrated that FPC-based wiring and analog circuits survive the thermal stresses occurring when cycling between room temperature and a cryogenic temperature of as low as 0.25 K (e.g. \cite{Holland2006SCUBA2,Lee2008POLARBEAR}). 
Attenuation from a polyimide substrate is also negligible if the substrate is thin enough, 
as demonstrated by the vacuum windows used in heterodyne receivers \cite{sugimoto2004}. 
These advantages motivated us toward the development of FPC-based metal mesh filters for mm/submm observational instruments. 

In this study, we demonstrate the possibility of producing mm/submm wideband BPFs using commercially available FPC fabrication technology. 
In Section~\ref{sec:design}, we describe the design of the metal mesh patterns. 
In Section~\ref{sec:simulation}, we detail an electromagnetic field simulation conducted to calculate the transmittance of the metal meshes. 
In Section~\ref{sec:fabrication}, we describe the fabrication of the FPC-based metal mesh filters. 
In Section~\ref{sec:Transmittance-Measurements}, we present and compare the results of the transmittance measurements with those of the simulations. 
Finally, we provide some concluding remarks in Section~\ref{sec:conclusions}.

\section{Metal Mesh Design}
\label{sec:design}

Typical metal mesh filters consist of inductive or capacitive periodic patterns \cite{ULRICH196737}. 
A BPF is realized by combining metal mesh patterns as a resonator. 
We designed a wideband BPF by stacking multiple layers of metal mesh filters possessing low-Q resonance patterns to achieve wide fractional bandwidth. 
Transmission properties of stacked metal mesh filters are often modeled by cascaded resonators in transmission line theory (e.g. \cite{wu2005multi}). 
The superposition of transmitted and reflected waves causes one or more resonances at certain frequencies, depending on the distances between layers. 
By choosing low-Q resonators for each layer, the resonance transmission peaks will overlap with each other such that a wide passband is achieved. 
At the same time, the change in the total phase shift due to resonators will be accumulated as the number of layers was increased, which leads to the steepening of the cut-off edges, in analogy to Chebyshev filters. 
However, when the number of layers is fixed, there is a trade-off between the wide fractional bandwidth and steepness of the cut-off edges. 
Here we put weight on wide fractional bandwidth so that we began with designing low-Q resonators for each layer. 

The Q-factor of a resonator depends on its geometry. 
A closely packed loop structure may have an advantage in such low-Q resonators; for example, a reflection with a fractional bandwidth of $\sim 1$ for a hexagonal loop was reported \cite{munk2005frequency}. 
Thus, a loaded grid, which is a complementary structure of the loop elements, is suitable for our requirements of a wideband BPF and we therefore adopted a loaded grid metal mesh structure.

\begin{center}
\begin{threeparttable}[ht]
\caption{Comparison between hexagonal and square inductive grids}
\label{table_lattices}
\begin{tabular}{ccc} \hline
    Lattice type & Hexagonal & Square \\ \hline\hline
    Geometry & 
    \begin{tabular}{c}
        \includegraphics[width=0.3\linewidth]{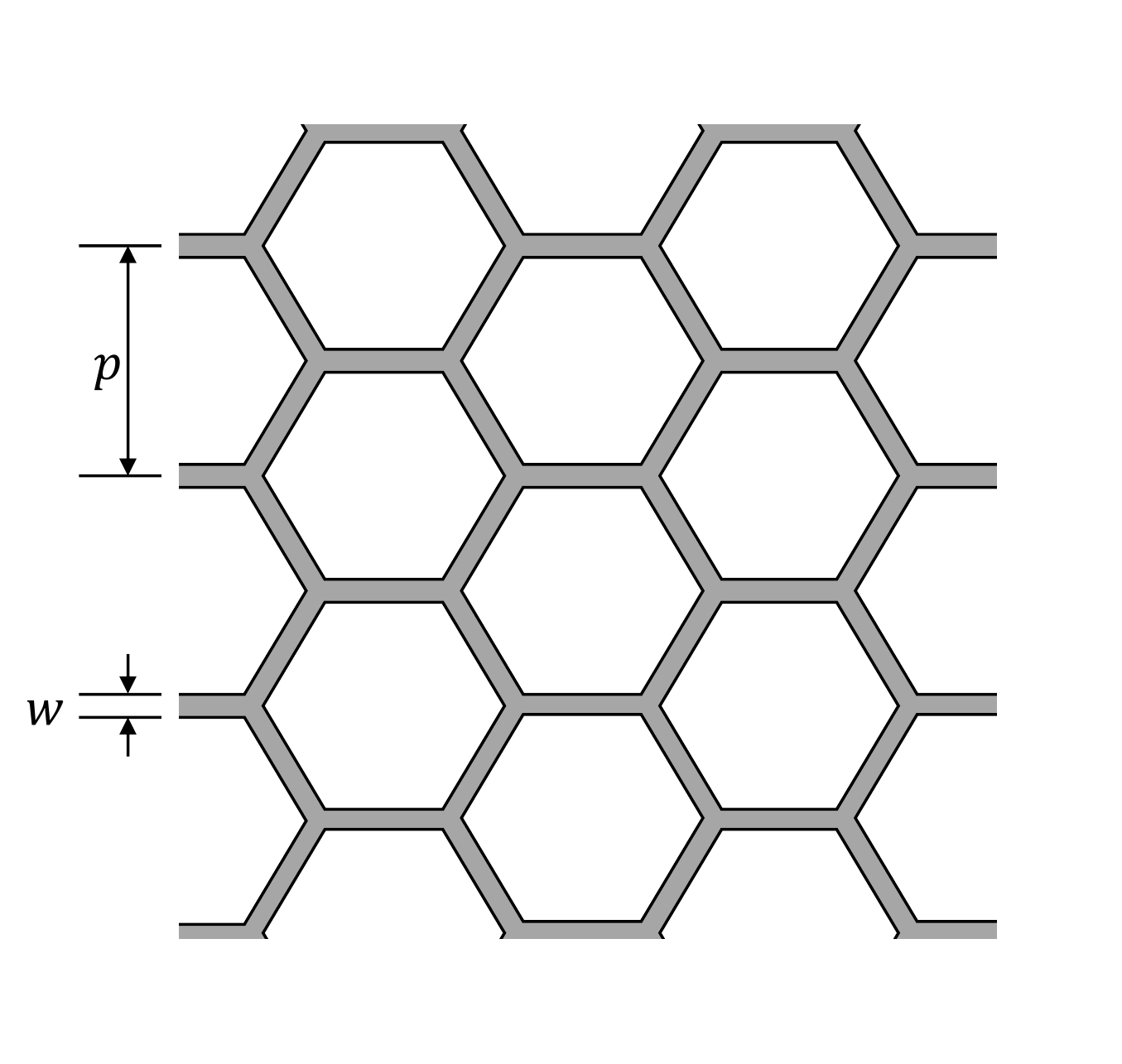}
    \end{tabular}
     & 
    \begin{tabular}{c}
        \includegraphics[width=0.3\linewidth]{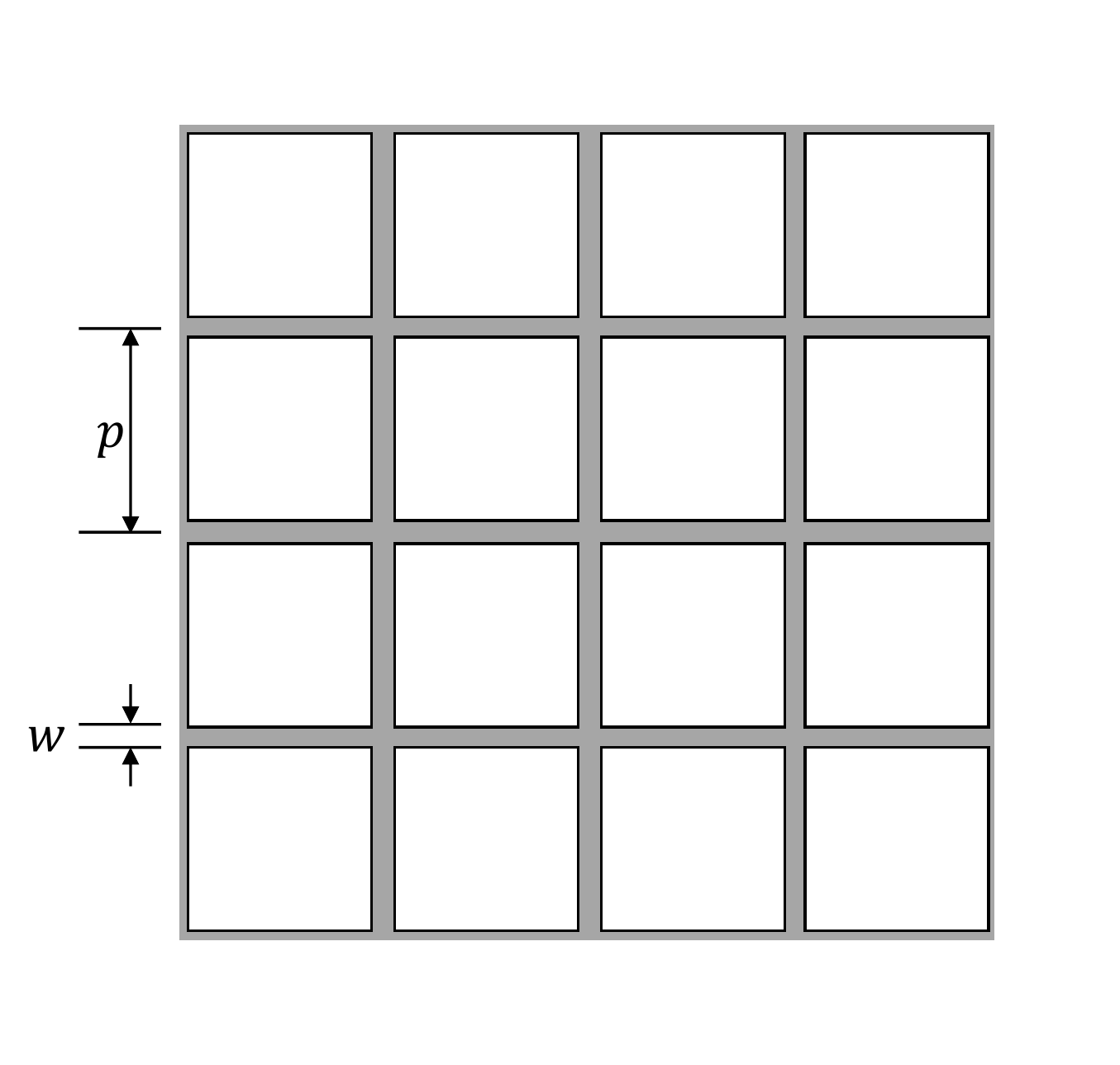} 
    \end{tabular} \\ \hline
    $\lambda_{\mathrm{diff}}$\tnote{a} \cite{munk2005frequency} & $\frac{\sqrt{3}}{2}p$ & $p$ \\ \hline
    $\lambda_{\mathrm{WG}}$\tnote{b} & $1.563\times\frac{2}{\sqrt{3}}(p-w)$ \cite{Komarov2011} & $2(p-w)$ \\ 
    $\displaystyle \frac{f_{\mathrm{diff}}}{f_{\mathrm{WG}}}$ & $\displaystyle \frac{1.563\times\frac{2}{\sqrt{3}}(p-w)}{\frac{\sqrt{3}}{2}p} = 2.084\frac{\lambda_\mathrm{diff}-\frac{\sqrt{3}}{2}w}{\lambda_\mathrm{diff}}$ & $\displaystyle \frac{2(p-w)}{p} = 2\frac{\lambda_\mathrm{diff}-w}{\lambda_\mathrm{diff}}$ \\ \hline
\end{tabular}
\begin{tablenotes}
\item[a] $\lambda_{\mathrm{diff}}$ denotes the diffraction-limit wavelength.
\item[b] $\lambda_{\mathrm{WG}}$ denotes the waveguide cut-off wavelength of a single aperture.
\end{tablenotes}
\end{threeparttable}
\end{center}

\begin{figure}[ht]
\centering
\includegraphics[width=0.8\linewidth]{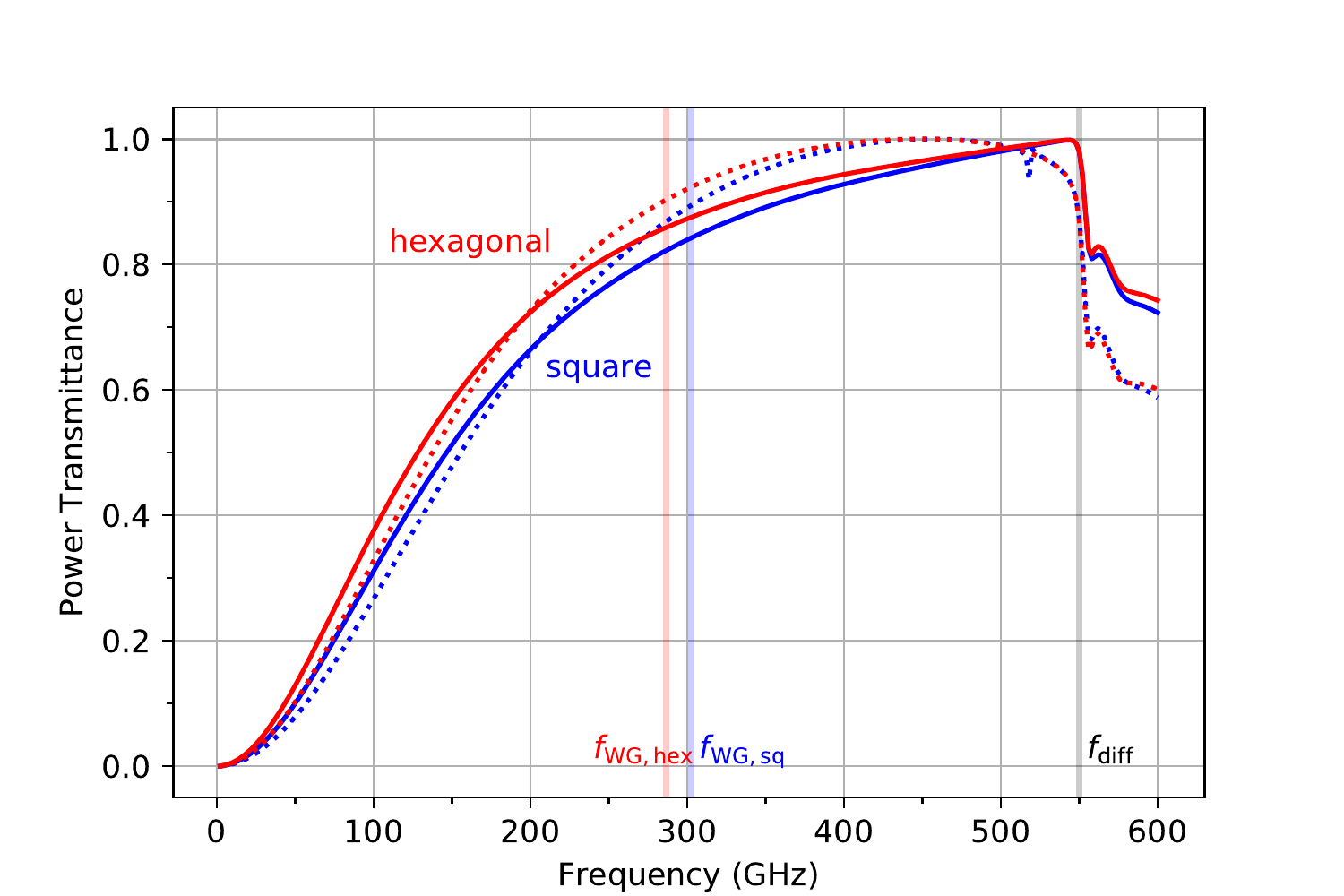}
\caption{Calculated transmittance of square and hexagonal grids. There are two types of simulated models in which the thicknesses of the metal and dielectric ($9~\mu\mathrm{m}$ and $12.5~\mu\mathrm{m}$, respectively) are considered ({\it dotted lines}) and ignored ({\it solid lines}). The patterns have a common diffraction-limit frequency (550~GHz) and metal linewidth ($w=50~\mathrm{\mu m}$). This figure shows that the passband of the hexagonal grid covers lower frequency and has potential to realize wider BPF than that of the square grid for a fixed diffraction-limit frequency and metal linewidth.}
\label{fig_sq_hex}
\end{figure}

A loaded grid is a combination of an inductive grid and capacitive loadings. 
An inductive grid can be considered as a periodic array of short waveguides. Compared with a waveguide, an inductive grid has a high-pass cut-off response, depending on the aperture size and shape. 
A comparison between hexagonal and square lattices is given in Table \ref{table_lattices}. 
The geometries of these inductive grids are characterized by the pitch $p$ and linewidth $w$. 
The larger the aperture is, the lower its waveguide (high-pass) cut-off frequency $f_{\mathrm{WG}}$. 
By contrast, its periodic structure causes a grating onset and diffraction loss above the diffraction-limit frequency $f_{\mathrm{diff}}$, which is in inverse proportion to the periodic length, defining the upper limit of the passband. 
Thus, our strategy for developing a wideband BPF is to increase the size of $f_{\mathrm{diff}}/f_{\mathrm{WG}}$, which means enlarging $(p-w)/p$, or to achieve $w\ll\lambda_{\mathrm{diff}}$, where $\lambda_{\mathrm{diff}}$ is the diffraction-limit wavelength. 
The recent fabrication precision of the FPC linewidth ($w=50~\mathrm{\mu m}$) meets our requirement for a narrow wire within the mm/submm wavelength ($\sim 300~\mathrm{GHz}$). 

To achieve a lattice with a large aperture relative to the periodic length when a finite narrow linewidth is given, we chose regular hexagonal and square lattices, which depend less on the polarization \cite{schuster2005micro}, and compared them as single layers of an inductive grid. 
We calculated the transmittance using COMSOL (see Section~\ref{sec:simulation}) and compared their high-pass cut-off characteristics with a fixed diffraction-limit frequency $f_{\mathrm{diff}}=550~\mathrm{GHz}$ for hexagonal and square grids (Fig. \ref{fig_sq_hex}). 
The pitches of the inductive metal mesh structure are $p=630~\mathrm{\mu m}$ and $546~\mathrm{\mu m}$ for hexagonal and square metal mesh structures, respectively. 
Fig. \ref{fig_sq_hex} shows that the passband of the hexagonal pattern covers lower frequency relative to $f_{\mathrm{diff}}$ than that of the square pattern. 
This result can be deduced from an analogy of $f_\mathrm{WG}$. 
The formulae of the waveguide cut-off wavelength $\lambda_\mathrm{WG}$ of hexagonal and square apertures are given in Table~\ref{table_lattices}. 
Although there is no exact analytical solution to a hexagonal waveguide mode, the normalized wavelength of the lowest eigenmode has been simulated numerically \cite{Komarov2011}, and we therefore apply its value here. 
If $f_\mathrm{diff}$ and a finite linewidth $w$ are given, the hexagonal aperture has a lower waveguide cut-off frequency than a square aperture.
Assuming $w=50~\mathrm{\mu m}$, the fractional bandwidth roughly calculated from $f_\mathrm{WG}$ and $f_{\mathrm{diff}}$ ($\lesssim 300~\mathrm{GHz}$) of a single-layer inductive grid reaches 0.6, which meets our requirement for a low Q-factor combined with capacitive loadings. 
Thus, we adopted a hexagonal loaded grid design for our basic wideband BPF.

As mentioned above, multiple-layer stacked filters with appropriate distances ($\sim$ a wavelength) can achieve a steep low- and high-pass cut-off frequency response. 
Thus, we demonstrate the use of single- and triple-layer stacked mesh filters in this study.

\section{Simulations}
\label{sec:simulation}

\subsection{Configuration details}

\begin{figure}[ht]
    \centering
    \includegraphics[width=0.8\linewidth]{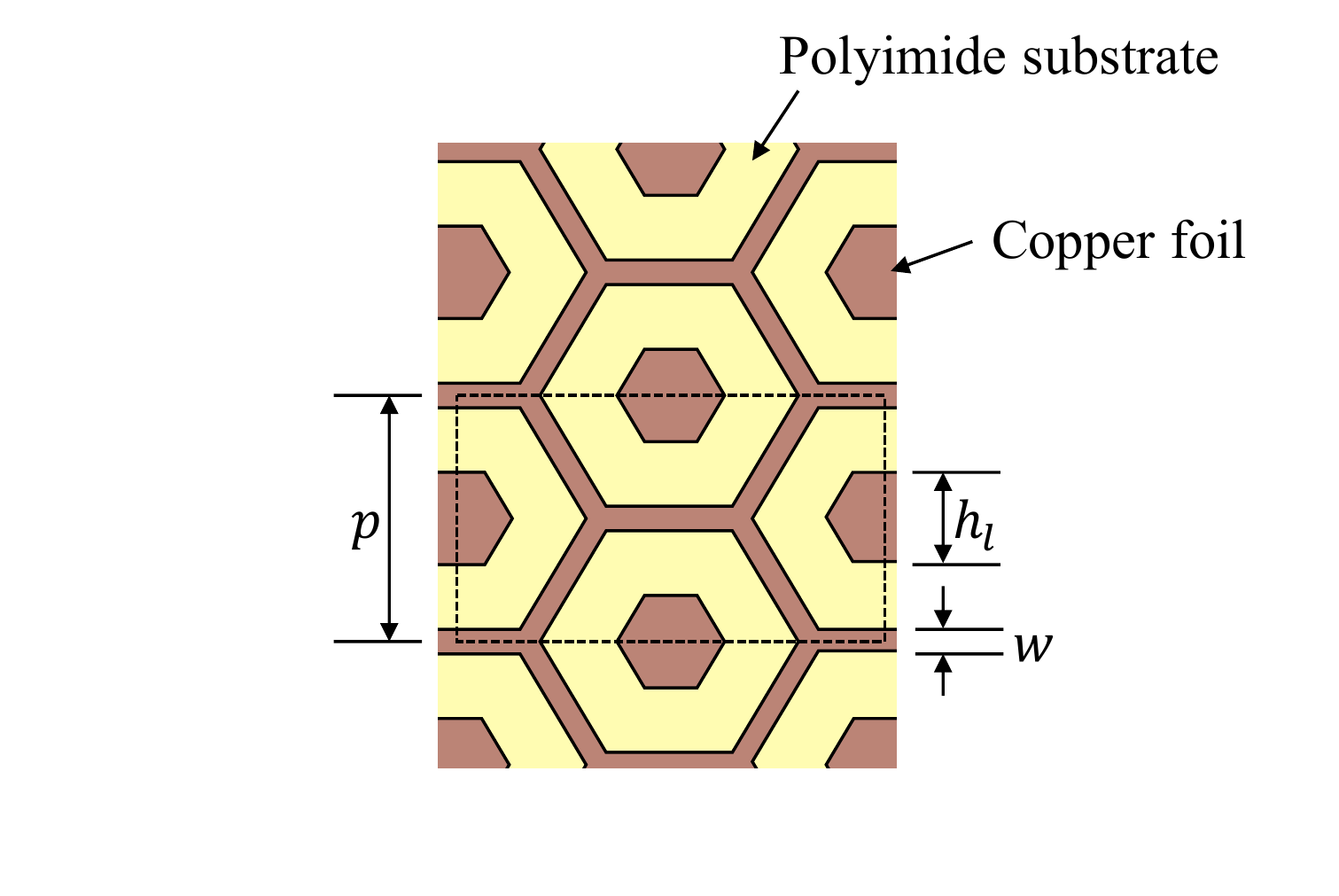}
    \caption{Geometry and notation of loaded hexagonal grid. The dashed rectangle represents an unit cell in the simulations. A $9~\mathrm{\mu m}$-thick copper pattern is printed on $12.5~\mathrm{\mu m}$-thick polyimide substrate.}
    \label{hexap_islands}
\end{figure}

The transmittances of FPC filters are calculated using the commercial software COMSOL Multiphysics and its RF module \cite{comsol}, which supports a three-dimensional electromagnetic field analysis using a finite element method. 
The notations of our model parameters (i.e., pitch $p$, loading height $h_l$, and linewidth $w$) are shown in Fig. \ref{hexap_islands}. 

Prototypes of the FPC were fabricated using a printed circuit board company, K2.inc \cite{K2inc}, the standard minimum process rule of which provides a copper pattern with a minimum linewidth and gaps of $50~\mu\mathrm{m}$ and $50~\mu\mathrm{m}$. 
The thicknesses of the copper and polyimide layer provided by this company are $9~\mu\mathrm{m}$ and $12.5~\mu\mathrm{m}$, respectively. 
These specifications are nearly ideal for a small dielectric absorption owing to the use of a thin substrate and small dependency on the incident angle from a thin metal pattern. 
In our simulation models, the thicknesses of the metal and substrate layers are therefore defined as $9~\mu\mathrm{m}$ and $12.5~\mu\mathrm{m}$, respectively. 

In our simulations, we ignored the dielectric loss by the metal pattern layer by assuming it as a perfect electric conductor (PEC). 
This is because the resistivity of copper is no more than $1.7\times10^{-8}~\Omega\cdot\mathrm{m}$ at room temperature, and the residual-resistance ratio (RRR) of copper is on the order of 10, and therefore the resistivity decreases one order of magnitude at an ultra-low temperature (0.25~K). 
In fact, if we allow for the resistivity of copper at room temperature instead of PEC in the simulations, the power loss is estimated to be less than 0.2\% for the loaded hexagonal grid.

The refractive index $n$ of the polyimide substrate was measured using THz-TDS at room temperature (see Section~\ref{sec:Transmittance-Measurements}), the result of which was $n=1.88$ and almost constant over the range of 0.1--1~THz. 
Thus, we adopted this value in our simulations. 
The loss-tangent $\tan\delta$ of polyimide was also obtained through our measurements at room temperature, and was $\tan\delta\lesssim0.02$ at 500~GHz. 
It was reported that polyimide has $\tan\delta=0.011$ at 150~GHz at room temperature, and within the range of 300--3,000~GHz $\tan\delta=0.030$ at room temperature and $\tan\delta=0.015$ at 5~K \cite{lau2006}. 
For a $12.5~\mu\mathrm{m}$ thickness, these absorption losses are expected to be less than 1\% in our observed bands (125--295~GHz), and therefore we ignore the dielectric absorption loss by the substrate during the simulations. 

\begin{figure}[ht]
    \begin{center}
        \subfigure[]{\includegraphics[width=0.4\linewidth]{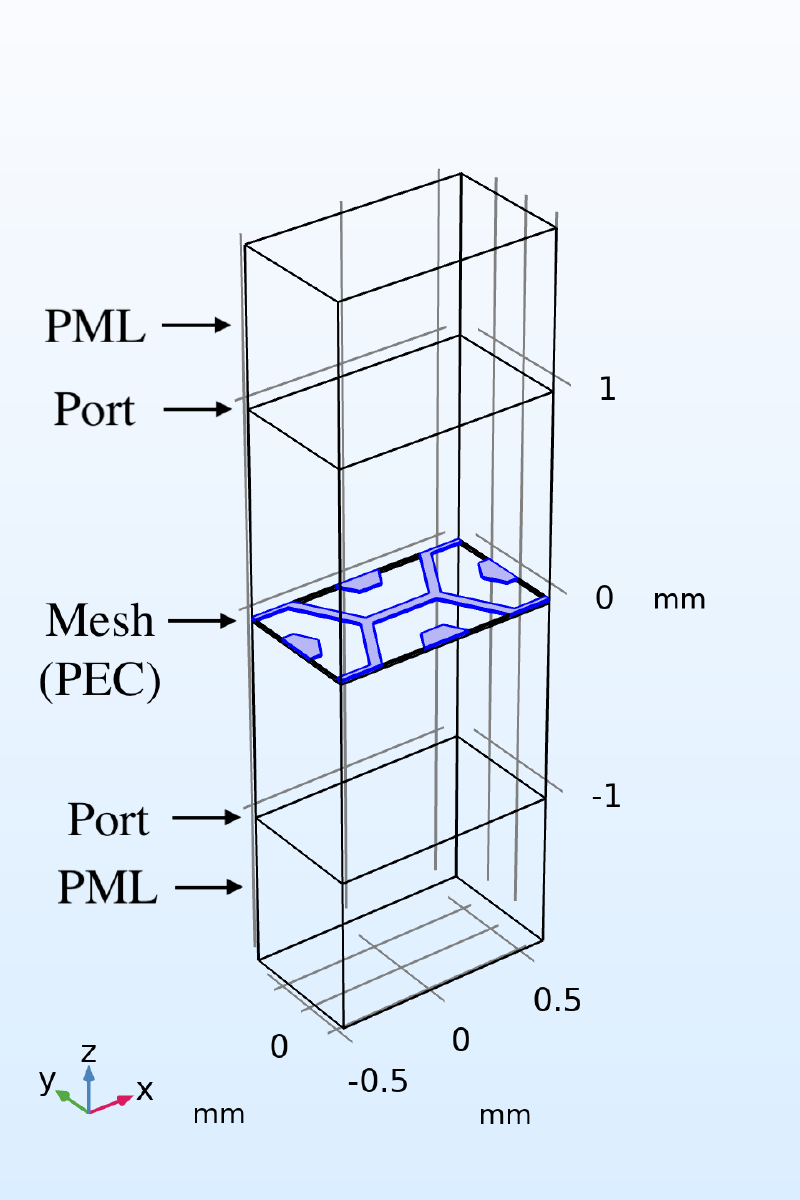}
        \label{comsol_model:left}}
        \subfigure[]{\includegraphics[width=0.4\linewidth]{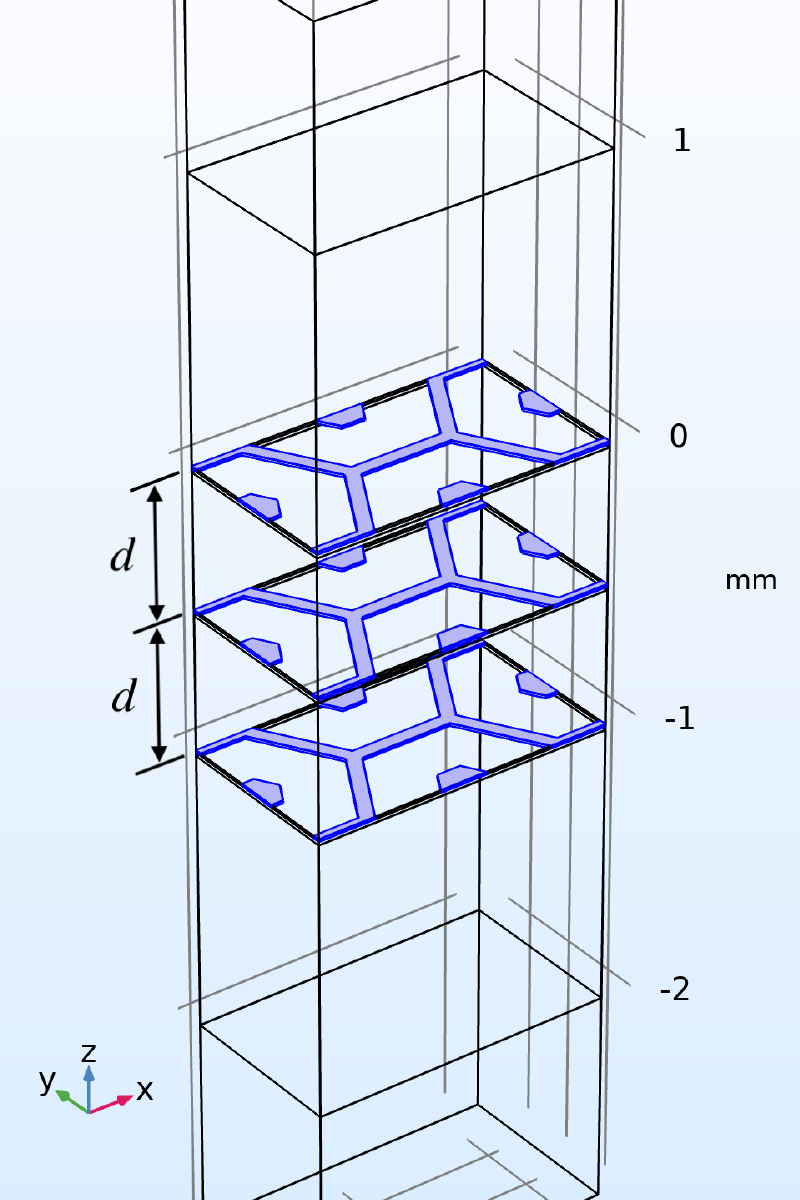}
        \label{comsol_model:right}}
        \caption{Simulated models of (a) loaded hexagonal grid and (b) three-layer-stacked mesh filter in COMSOL multiphysics/RF module. The blue domain is designated as a perfect electric conductor (PEC). Two perfectly matched layers (PMLs) are set outside of the ports.}
    \end{center}
\end{figure}

To reduce the computational cost for periodic patterns, we use the periodic boundary condition shown in Fig. \ref{comsol_model:left}. 
The unit cell of the mesh pattern is located in the middle of a box. 
Exciting and listener ports are designated on the top and bottom of the box, and two perfectly matched layers (PMLs) are set outside of the ports to avoid unwanted reflections. 
The incident beam is assumed to be a plane wave. 
Because we do not intend our filters to be tilted toward an incident beam, the transmittance is simulated at normal incidence.

In this study, we demonstrate the use of high-frequency scaled models in the evaluation of our required passband (i.e., 125--295~GHz) and its slope on the lower-frequency side through our measurement setup (0.1--1~THz, see Section~\ref{sec:Transmittance-Measurements}). 
These scaled models are also aimed at confirming the feasibility of a higher-frequency filter ($\sim$ 300--700~GHz). 
To reduce the experimental noise from the absorption by water vapors at near 550~GHz, the models have been scaled to achieve $f_{\mathrm{diff}}=550~\mathrm{GHz}$, which corresponds to $p=630~\mu\mathrm{m}$ for a hexagonal lattice. 
Three-layer-stacked filters and single meshes were simulated. 
In the simulations of the stacked filters, the meshes were set to have their copper surfaces face in the incident direction, aligned with each other, and separated with distances $d$ of roughly one half of the wavelength, as shown in Fig. \ref{comsol_model:right}.

\subsection{Parameter dependencies of transmittance}

Herein, we investigate the parameter dependencies of the transmittance for our filter design.

\begin{figure}[ht]
    \centering
    \includegraphics[width=\linewidth]{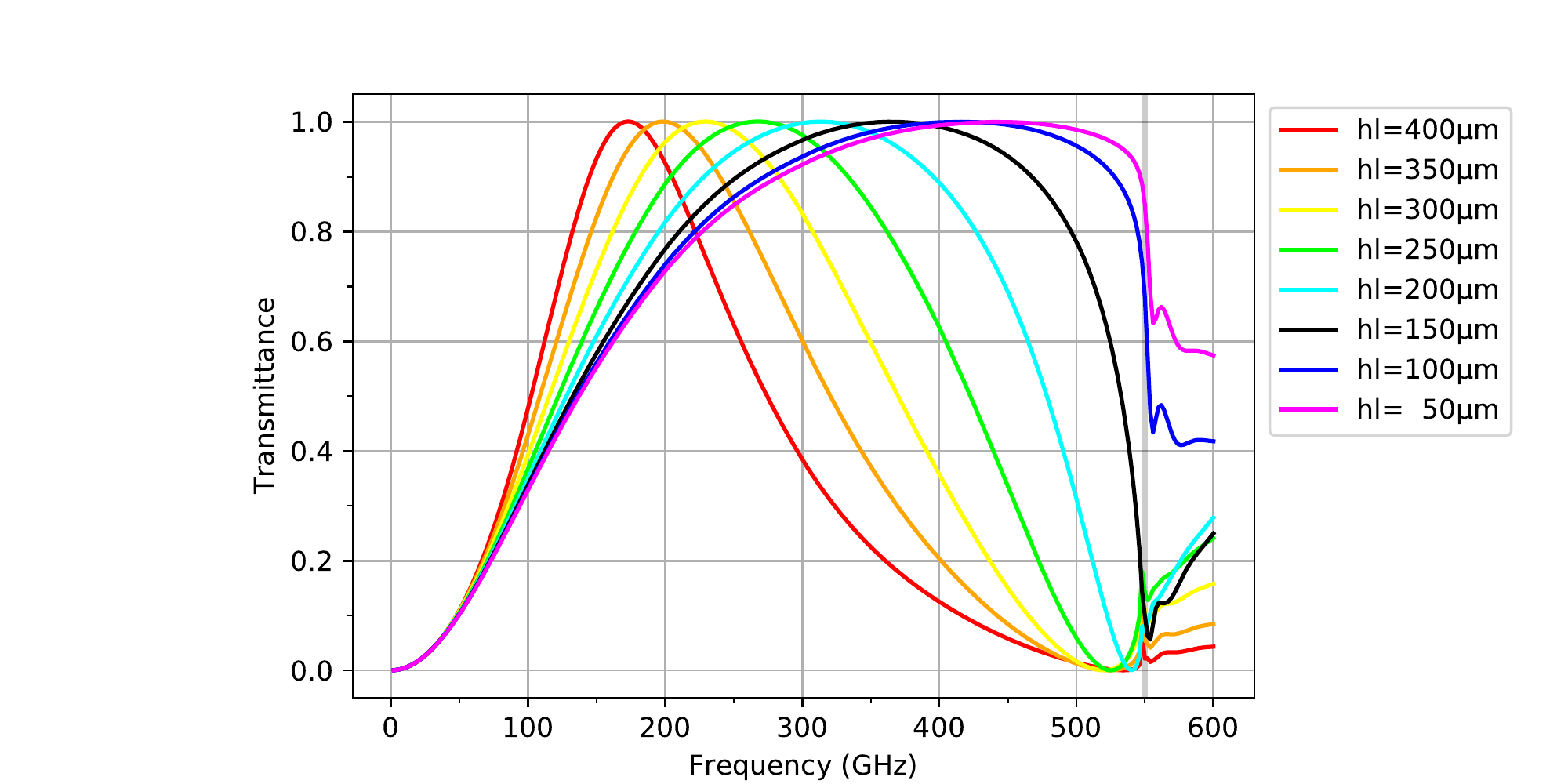}
    \caption{Simulated loading height $h_l$ dependence of a single resonant mesh. The other parameters are fixed at $p=630~\mathrm{\mu m}$ and $w=50~\mathrm{\mu m}$. The diffraction-limit frequency is 550~GHz.}
    \label{sim_hl}
\end{figure}

First, we simulated the transmittances for different loading heights $h_l$, as shown in Fig. \ref{sim_hl}. 
The linewidth $w$ was fixed to $50~\mu\mathrm{m}$. 
As $h_l$ increases, the peak frequency decreases and the bandwidth becomes monotonically narrower.
Based on the definitions of the peak frequency and fractional bandwidth of the equivalent parallel circuit, this dependency is interpreted as an increase in the capacitance with the loading height, particularly the separation between the loading and inductive grid. 
Hereafter, we use $h_l= 150~\mu\mathrm{m}$ and $w=50~\mu\mathrm{m}$ for a low-Q resonator.

\begin{figure}[ht]
    \centering
    \includegraphics[width=\linewidth]{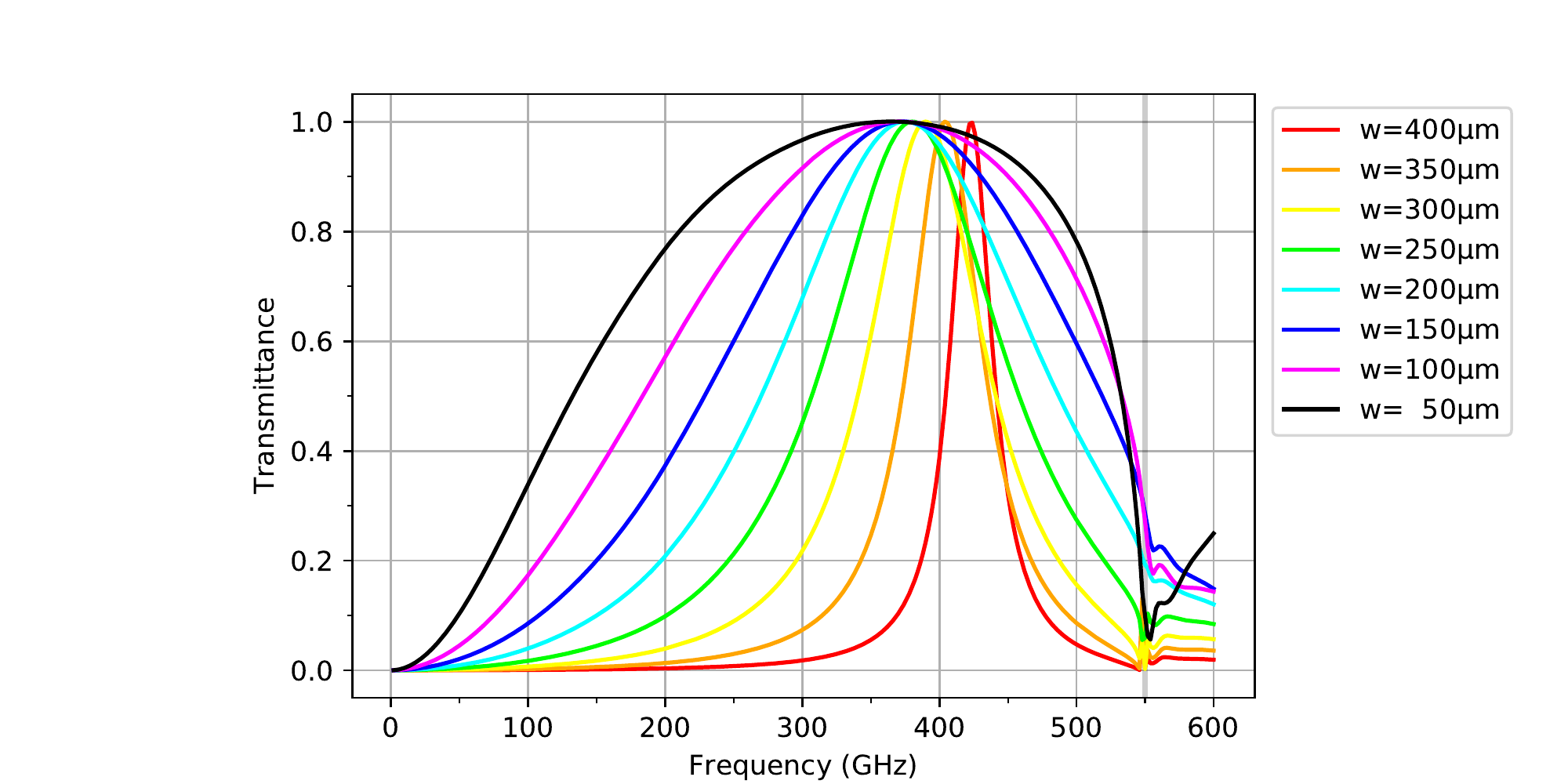}
    \caption{Simulated linewidth $w$ dependence of a single resonant mesh. The other parameters are fixed at $p=630~\mathrm{\mu m}$ and $h_l=150~\mathrm{\mu m}$. The diffraction-limit frequency is 550~GHz.}
    \label{sim_w}
\end{figure}

Second, the simulated transmittances for different linewidths $w$ are shown in Fig. \ref{sim_w}. 
As $w$ increases, the bandwidth becomes narrower, whereas the peak frequency remains nearly constant. 
We can assume that this dependency is caused by two factors: (i) an increase in capacitance with a narrower separation between the loading and inductive grid, and (ii) a decrease in the inductance with a wider grid. 

\begin{figure}[ht]
    \centering
    \includegraphics[width=\linewidth]{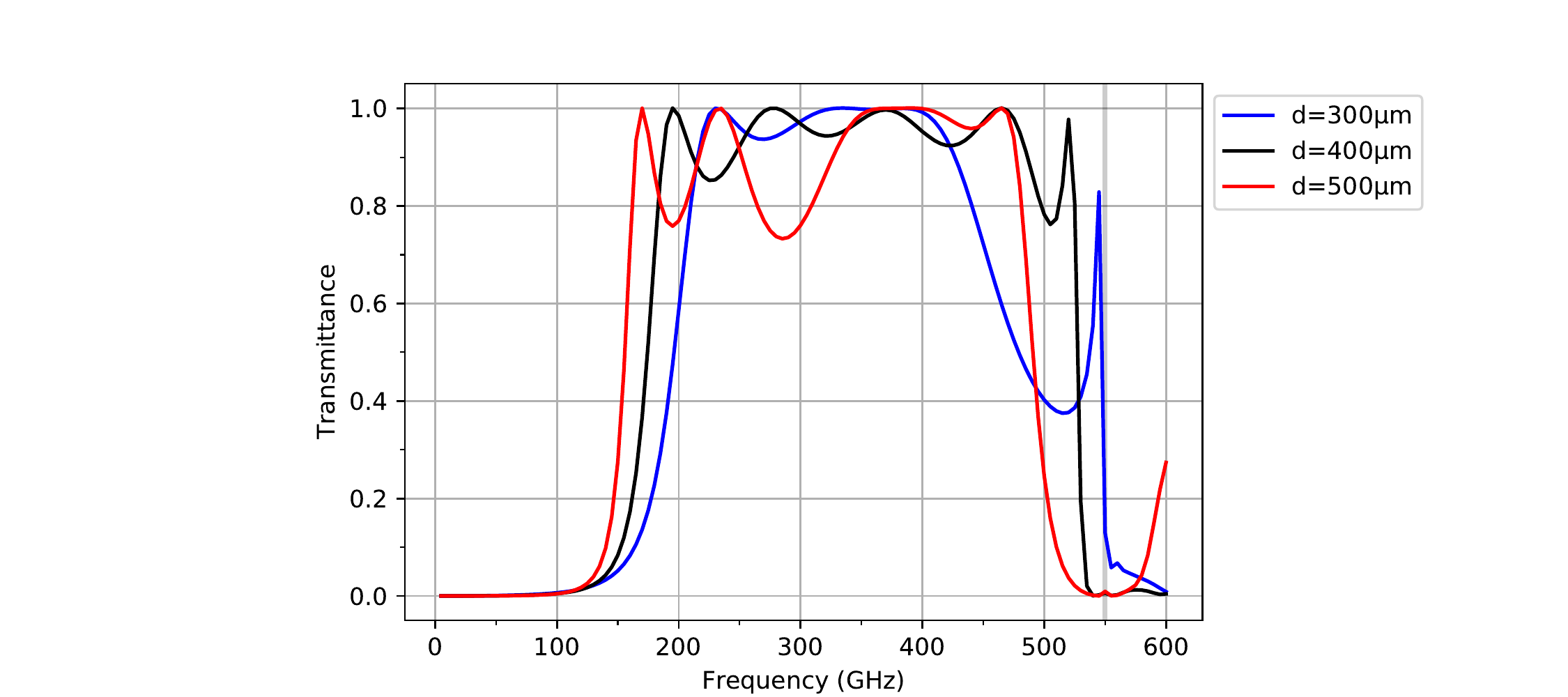}
    \caption{Simulated distance $d$ dependence of three-layer-stacked mesh. The other parameters are fixed at $p=630~\mathrm{\mu m}$, $h_l=150~\mathrm{\mu m}$ and $w=50~\mathrm{\mu m}$. The diffraction-limit frequency is 550~GHz. The passband of $d= 400~\mathrm{\mu m}$ matches our application owing to its relatively steep cut-off, small ripples and wide bandwidth.}
    \label{sim_3lyr}
\end{figure}

Third, simulated transmittances of three-layer-stacked filters for different distances $d$ are as shown in Fig. \ref{sim_3lyr}. 
Ripples occur in the passband depending on the stacking configuration. 
As $d$ increases, the high- and low-pass cut-off becomes steep with a variation in the number and depths of the ripples. 
As seen in a general multi-element LC filter circuit design, we can also see the same trade-off between the cut-off and ripples. 
We adopted $d= 400~\mu\mathrm{m}$ along with $h_l= 150~\mu\mathrm{m}$ and $w=50~\mu\mathrm{m}$ owing to its small ripples and wide bandwidth. 

The above simulation results show that under the geometrical constraint, the bandwidth and center frequency of the loaded hexagonal mesh filters can be controlled based on the linewidth and loading height.
Moreover, we see a qualitative relationship between the equivalent circuit and geometrical parameters (i.e., loading height $h_l$ and linewidth $w$). 
This suggests that the stacking of metal meshes is equal to coupling parallel resonators in an equivalent circuit when assembling a Chebyshev-like BPF.

Note that the thicknesses of the dielectric and metal layers must not be ignored during the simulation. 
The metal layer coupled to the dielectric layer decreases the pass-band frequency.
By contrast, the thickness of the metal layer is expected to be related to the waveguide cut-off characteristics, which steepens the cut-off.

\section{Fabrication}
\label{sec:fabrication}

\begin{figure}[ht]
\centering
\includegraphics[width=0.5\linewidth]{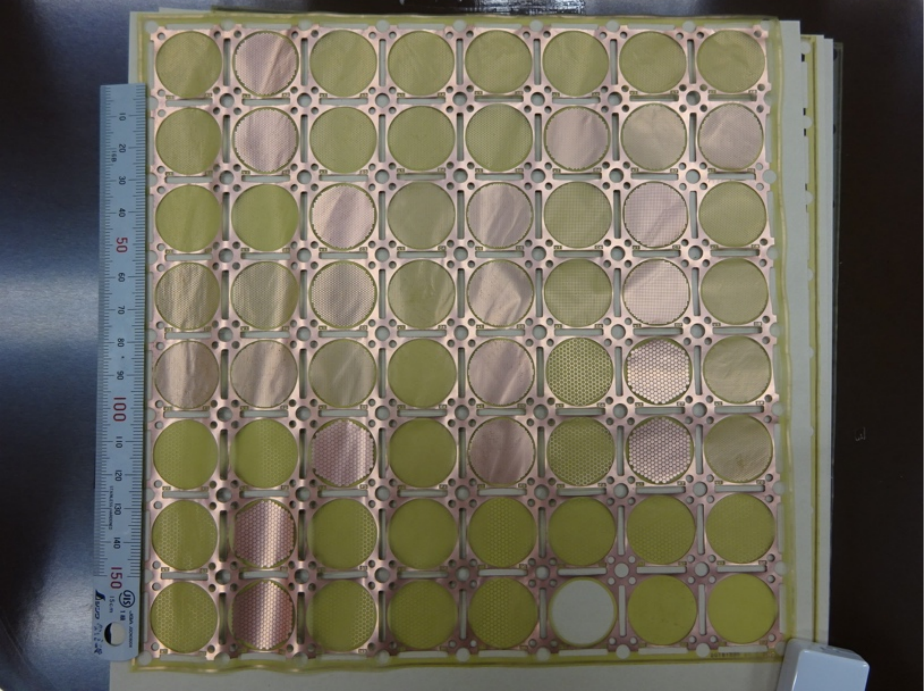}
\caption{An example of a FPC sheet containing sixty-four filter patches fabricated through a commercial process.}
\label{fig_prototype}
\end{figure}

\begin{figure}[ht]
    \begin{center}
        \subfigure[]{\includegraphics[width=0.4\linewidth]{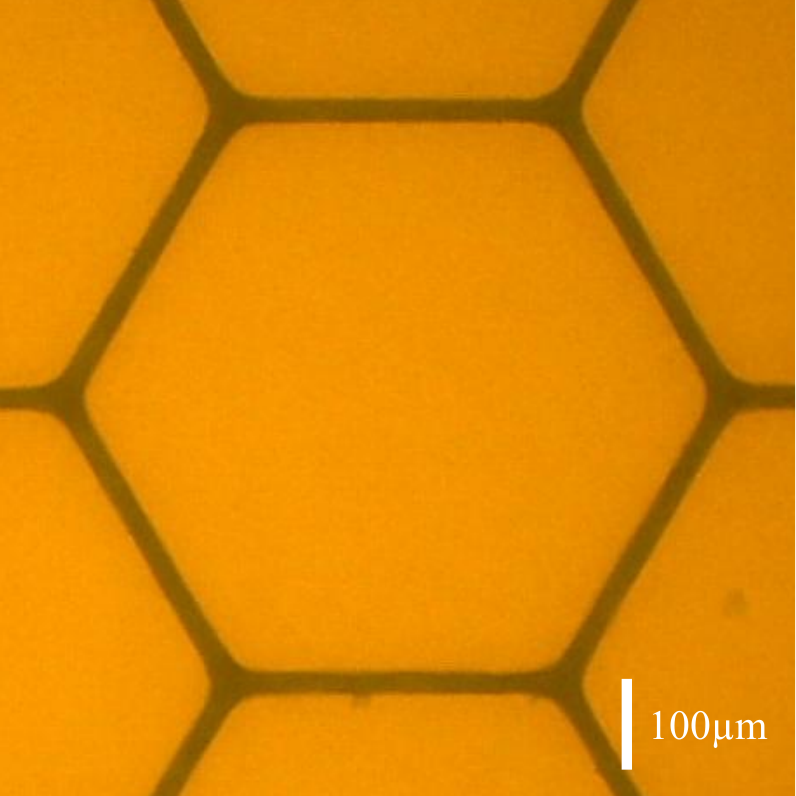}
        \label{microscope:left}}
        \subfigure[]{\includegraphics[width=0.4\linewidth]{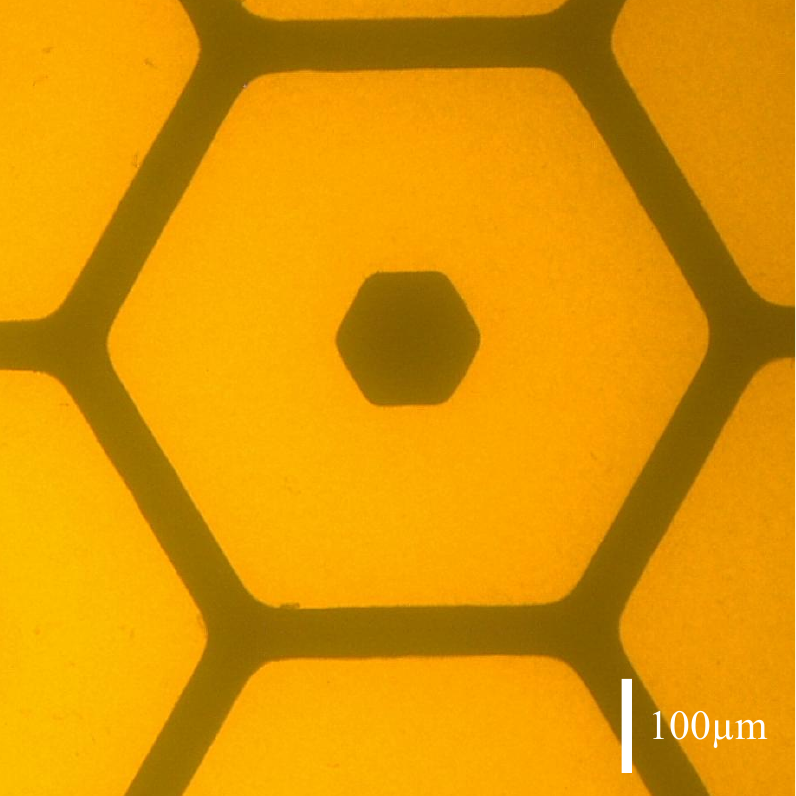}
        \label{microscope:right}}
        \caption{Microscopic images of (a) Model~1: a hexagonal inductive grid and (b) Model~2: a loaded hexagonal grid. The dark part is the copper pattern.}
        \label{microscope}
    \end{center}
\end{figure}

We created sixty-four 18-mm diameter (which is larger than the beam waist of our measurement setup at higher than $0.1~\mathrm{THz}$) filter patches of various design parameters on a 20~cm $\times$ 20~cm sheet.

A prototype FPC sheet is shown in Fig. \ref{fig_prototype}. 
The FPC sheets tend to roll by weakening the sheet and unbalancing the tension of the copper foil due to the holes on each filter patch used for alignment, in addition to the fabrication of numerous patches on a single sheet. 
In a future fabrication, each filter will be larger and have a smaller number of holes than the prototype. 
Even for the present prototype, such rolling does not worsen the surface roughness and therefore the transmittance is not affected if the sheet is stretched with support rings and spacers. 

We conducted a microscopic inspection of the fabricated patterns. 
Two types of observed hexagonal copper structures are shown in Fig. \ref{microscope}. 
We confirmed that the measured pitches $p$ are consistent within an accuracy of $<10~\mathrm{\mu m}$ for all patterns.
By contrast, the linewidths $w$ and loading heights $h_l$ showed a $\sim25~\mathrm{\mu m}$ smaller than the designed value, which was uniform in each sheet and among different sheets.
These results can be explained by an over-etching of the metal structures through the photo-lithography process, which will not occur in the future because the fabrication accuracy can be improved by defining the etching conditions in advance.

\section{Transmittance Measurements}
\label{sec:Transmittance-Measurements}

\begin{table*}[ht]
    \centering
    \caption{Specifications of the THz-TDS}
    \label{THz-TDS}
    \begin{tabular}{ll} \hline
        Name of instrument: & \\
        \hspace{5mm} Source & Advantest TAS1110 \\
        \hspace{5mm} Detector & Advantest TAS1230 \\ 
        \hspace{5mm} Sampling analysis system & Advantest TAS7500TS \\ \hline
        Measurable frequency range & 0.1--4~THz \\
        Frequency resolution & 3.8~GHz \\
        Focal ratio of optics & $\sim 4$ at target sample \\
        $1/f$ knee frequency & 3~mHz (30~sec interleaving to mitigate) \\ \hline
    \end{tabular}
\end{table*}

We measured the transmittance of our prototypes using a terahertz time-domain spectroscopy (THz-TDS). 
The specifications of the THz-TDS are shown in Table~\ref{THz-TDS}. 
The THz wave source of the THz-TDS generates linearly polarized pulses and covers a range of 0.1--4~THz.
Target samples and the background were switched every 30~s to remove the low-frequency fluctuations of the measurement system, and the reproducibility of the transmittance measurement reached less than 0.5\%.
The frequency scale was calibrated using intense absorption lines of water vapor in the atmosphere as the frequency standards. 
Target samples were set perpendicular to the optical axis of the measurement system. 
To check the different responses between the two polarizations, each sample was measured at two angles, i.e., aligning the mesh along the polarization axis and rotating the mesh by $90^\circ$.

\begin{figure}[ht]
    \centering
    \includegraphics[width=\linewidth]{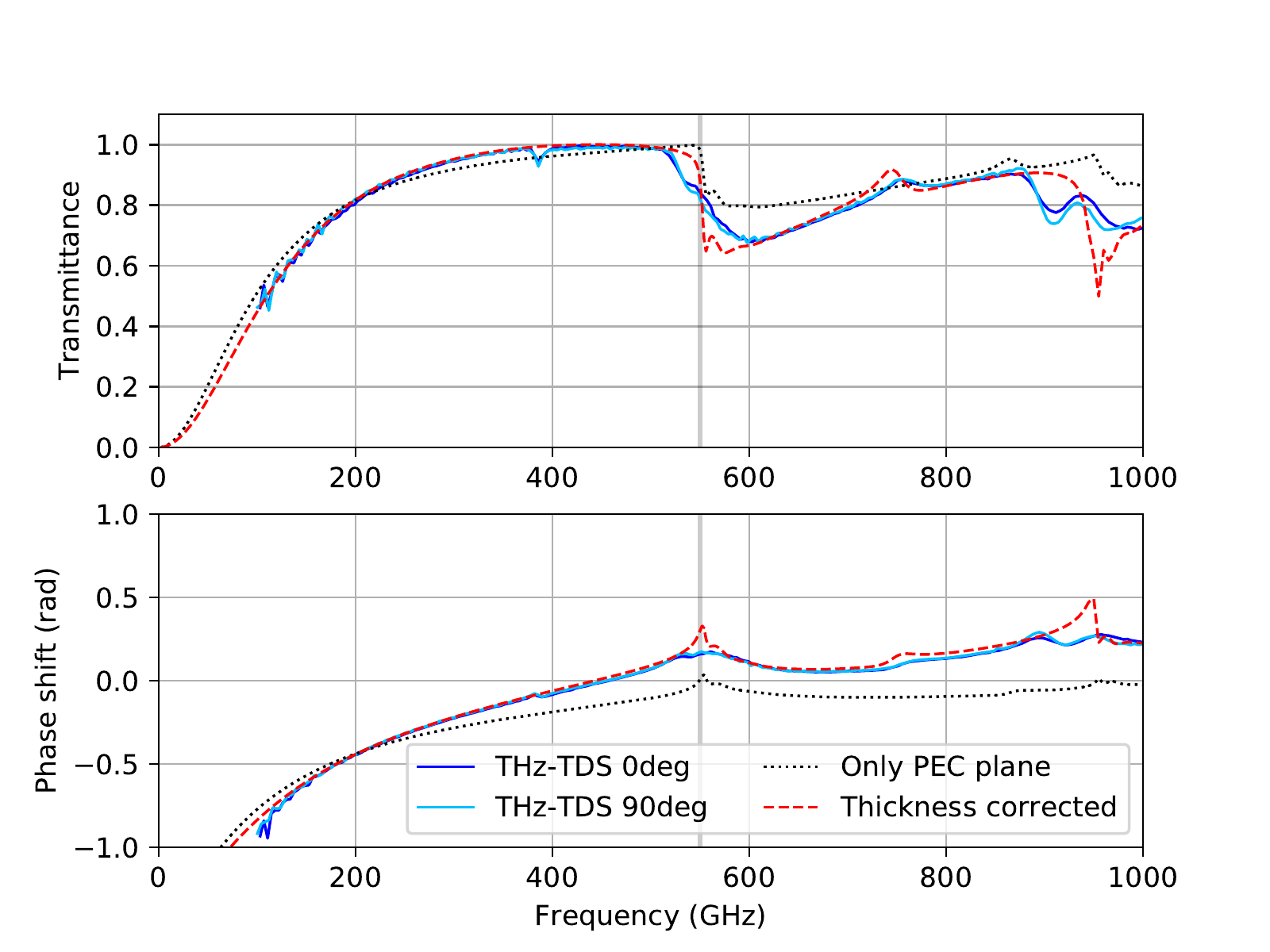}
    \caption{The transmittance ({\it top}) and phase shift ({\it bottom}) of a hexagonal inductive grid (Model~1). Measured transmittance of orthogonal polarization angles (blue and cyan solid), simulation results considering the thickness of the metal and dielectric layers (red dash), and ignoring the thickness of metal and dielectric layers (black dot), are shown. The vertical line shows the diffraction-limit frequency. The parameters are $p=630~\mu\mathrm{m}$ and $w=27~\mu\mathrm{m}$. A high-pass performance is seen below the diffraction-limit frequency of 550~GHz.}
    \label{fig_rC1}
\end{figure}

We compared the transmittance characteristics from the measurements and simulations. 
First, the power transmittance and phase shift of a single-layer hexagonal high-pass grid (which we define herein as Model~1) are as shown in Fig. \ref{fig_rC1}. 
The measured transmittance characteristics of Model~1 are in good agreement with our simulations when considering the effect of the over-etching. 
As expected from the geometry and simulations, there is little difference in the transmittance for polarization angle.

\begin{figure}[ht]
    \centering
    \includegraphics[width=\linewidth]{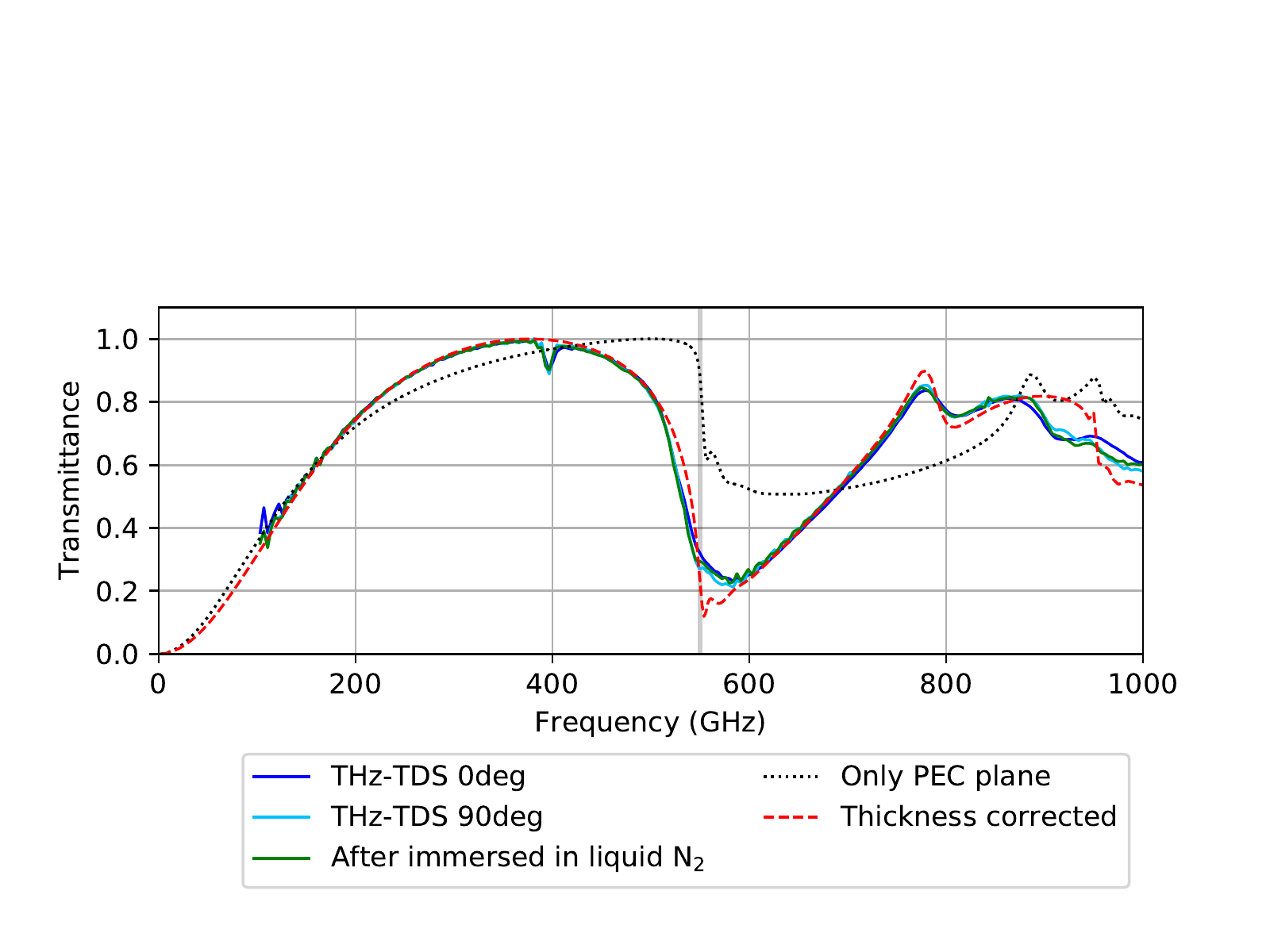}
    \caption{Transmittance of a hexagonal resonant pattern (Model~2) in measurements with orthogonal polarization angles and in simulation. The legend is the same as shown in Fig. \ref{fig_rC1}, except for the measured transmittance after immersion in liquid nitrogen (green solid). The parameters are $p=630~\mu\mathrm{m}, w=55~\mu\mathrm{m}$ and $h_l=141~\mu\mathrm{m}$. A band-pass performance is seen below the diffraction-limit frequency of 550~GHz.}
    \label{fig_dD8}
\end{figure}

Second, the power transmittance of a single-layer loaded hexagonal band-pass grid (which we define herein as Model~2) is as shown in Fig. \ref{fig_dD8}. 
As with Model~1, the measured transmittance of Model~2 is in good agreement with the simulations, and there is little difference in the transmittance for the two polarization angles. 

Because the loadings are not connected to the grid, they may be peeled off. 
To test the possible change in the optical property by the peeling-off of loading of the foil, or through the tension of the film during thermal cycling, we immersed Model~2 into liquid nitrogen for approximately 3 minutes twice and measured its transmittance at room temperature again. 
We found that the transmittance of Model~2 did not change, and a peeling-off of the loading did not occur. 

\begin{figure}[ht]
    \centering
    \includegraphics[width=\linewidth]{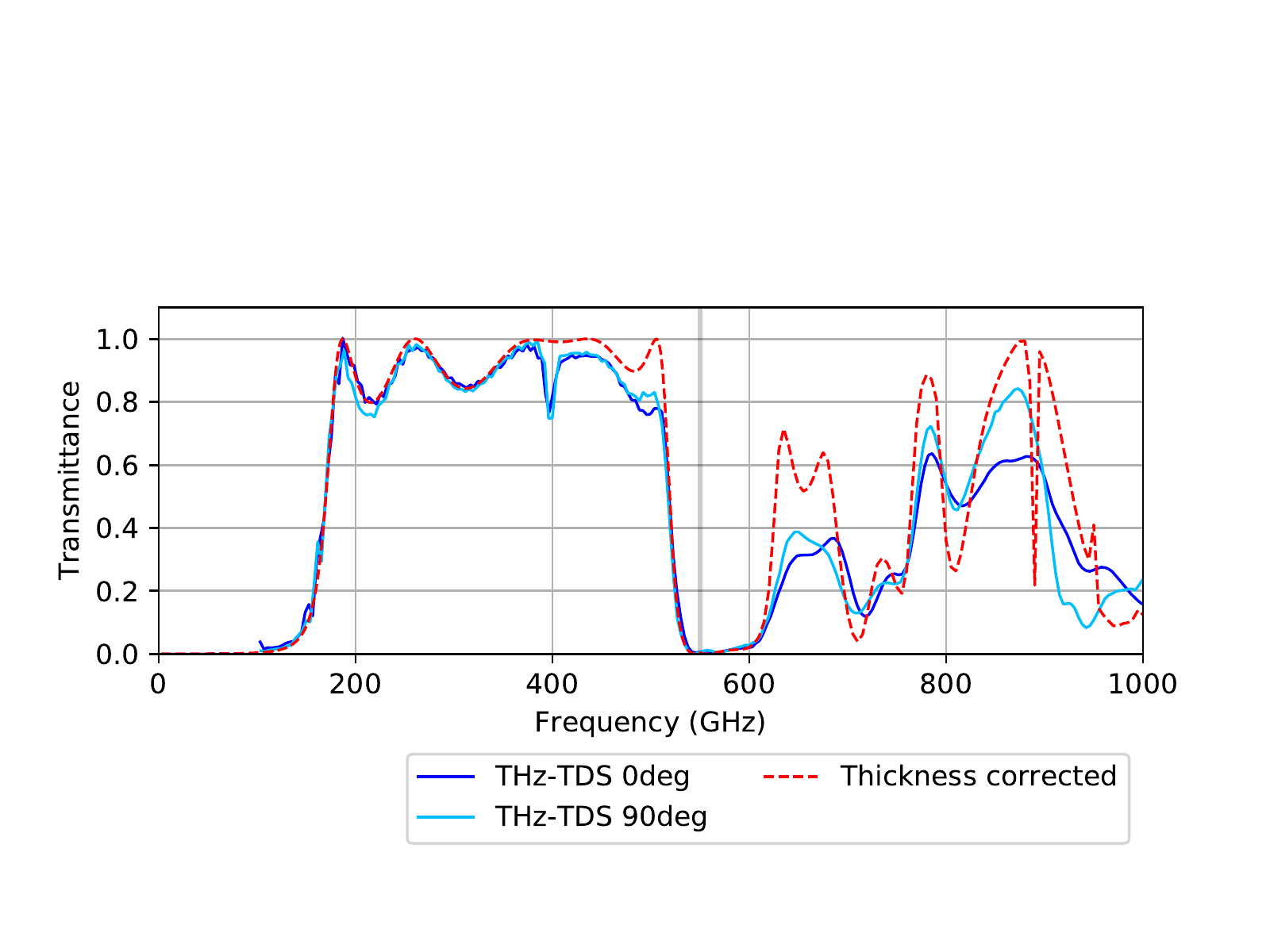}
    \caption{Transmittance of a three-layer-stacked filter of hexagonal resonant pattern (Model~2) in measurements with orthogonal polarization angles and in simulation. The legend is the same as shown in Fig. \ref{fig_rC1}. The parameters are $p=630~\mu\mathrm{m}, w=55~\mu\mathrm{m}, h_l=141~\mu\mathrm{m}$ and $d=440~\mu\mathrm{m}$. A band-pass performance with fractional bandwidth of 1.0 and a steep low- and high-pass cut-off is seen below the diffraction-limit frequency of 550~GHz. In the diffraction region a high-frequency leak occurs, which is roughly reproduced by the simulation.}
    \label{fig_dD8_3lyr}
\end{figure}

Finally, the transmittance of a BPF of the three-layer-stacked Model~2 is shown in Fig. \ref{fig_dD8_3lyr}. 
The three meshes were set to have their copper surfaces face in the incident direction, and were aligned and stacked together with spacers to achieve $d=440~\mu\mathrm{m}$. 
The 3~dB high-pass and low-pass cut-off frequencies of this stacked filter are 170 and 520~GHz, respectively. 
This fractional bandwidth of 1.0 is sufficiently wide for our purposes. 
Although there are ripples in the passband, which reaches 20~\% at maximum, the average transmittance in the passband is approximately 90~\% and meets our requirement.
By contrast, a high-frequency leak occurs above the diffraction-limit frequency. 
To attenuate this high frequency leak in practice, an additional low-pass metal mesh filter should be used. 

We can see a narrow resonance at $\sim$400~GHz within each transmission band. 
This resonance is known as Wood's anomaly \cite{ADE2006}, which is polarization-sensitive diffraction effect in which the transmittance strongly depends on the incident angle and polarization. 
The optics of a THz-TDS has a focal ratio of $\sim 4$ at the sample position, which means the incident beam includes an oblique component, which is expected to be a cause of Wood's anomaly.

\section{Conclusions}
\label{sec:conclusions}

We verified the feasibility of applying a commercial FPC fabrication process to a quasi-optical wideband BPF in the mm/submm range. 
By stacking three layers of metal mesh filters having a loaded hexagonal grid, a steep cut-off and large fractional bandwidth ($\sim 1$) were achieved, in the frequency range of 170--520~GHz.
Thus, we have shown that the commercial FPC fabrication technology can overcome the cost and turnaround time challenges to develop on-demand mm/submm metal mesh filters. 
Our next step is to fabricate a wideband BPF covering 125--295~GHz for the new multichroic continuum camera which enables to observe three frequency bands simultaneously when combined with on-chip frequency filters. 
In the future, further progress in FPC fabrication technology will give an access to larger area with finer patterns, which makes this technique a good candidate for realizing metal mesh filters for large aperture cameras and THz instruments.

\section*{Funding}

Japan Society for the Promotion of Science (JSPS) KAKENHI (17H02872, 17H06130, 19K14754); NAOJ Research Coordination Committee, NINS (1901-0102).

\section*{Acknowledgments}

S.U. is supported by FoPM, WINGS Program, the University of Tokyo.

\section*{Disclosures}

The authors declare no conflicts of interest.


\bibliography{FPCfilter_uno2020}

\end{document}